\documentclass[twocolumn]{aastex631}
\usepackage{amsmath}
\usepackage[T1]{fontenc}
\usepackage[utf8]{inputenc} 
\usepackage{upgreek}
\usepackage{breqn}
\usepackage{bm}
\DeclareSymbolFont{matha}{OML}{txmi}{m}{it}
\DeclareMathSymbol{\varv}{\mathord}{matha}{118}

\newcommand{\PSUAA}{Department of Astronomy \& Astrophysics, The Pennsylvania State University, 525 Davey Laboratory, 251 Pollock Road, University Park, PA, 16802, USA}
\newcommand{\PSUCEHW}{Center for Exoplanets and Habitable Worlds, The Pennsylvania State University, 525 Davey Laboratory, 251 Pollock Road, University Park, PA, 16802, USA}
\newcommand{\PSETI}{Penn State Extraterrestrial Intelligence Center, The Pennsylvania State University, 525 Davey Laboratory, 251 Pollock Road, University Park, PA, 16802, USA}
\newcommand{\PSUICDS}{Institute for Computational and Data Sciences, The Pennsylvania State University, University Park, PA, 16802, USA}
\newcommand{\UA}{Steward Observatory, The University of Arizona, 933 N.\ Cherry Ave, Tucson, AZ 85721, USA}

\newcommand{\Penn}{Department of Physics and Astronomy, University of Pennsylvania, 209 S 33rd St, Philadelphia, PA 19104, USA}

\newcommand{\GSFC}{NASA Goddard Space Flight Center, Greenbelt, MD 20771, USA}
\newcommand{\NOAO}{U.S. National Science Foundation National Optical-Infrared Astronomy Research Laboratory, 950 N.\ Cherry Ave., Tucson, AZ 85719, USA}

\newcommand{\Macquarie}{School of Mathematical and Physical Sciences, Macquarie University, Balaclava Road, North Ryde, NSW 2109, Australia}

\newcommand{\JPL}{Jet Propulsion Laboratory, California Institute of Technology, 4800 Oak Grove Drive, Pasadena, California 91109}

\newcommand{\UCI}{Department of Physics \& Astronomy, The University of California, Irvine, Irvine, CA 92697, USA}

\newcommand{\Carnegie}{Earth and Planets Laboratory, Carnegie Institution for Science, 5241 Broad Branch Road, NW, Washington, DC 20015, USA}
\newcommand{\PSUICS}{Institute for Computational and Data Sciences, The Pennsylvania State University, University Park, PA, 16802, USA}
\newcommand{\PSUCASt}{Center for Astrostatistics, 525 Davey Laboratory, The Pennsylvania State University, University Park, PA, 16802, USA}

\newcommand{\TIFR}{Department of Astronomy and Astrophysics, Tata Institute of Fundamental Research, Homi Bhabha Road, Colaba, Mumbai 400005, India}

\newcommand{\FlatironCCA}{Center for Computational Astrophysics, Flatiron Institute, 162 Fifth Avenue, New York, NY 10010, USA}
\newcommand{\UAm}{Anton Pannekoek Institute for Astronomy, 904 Science Park, University of Amsterdam, Amsterdam, 1098 XH}
\newcommand{\UIUC}{Department of Astronomy, University of Illinois at Urbana-Champaign, Urbana, IL 61801, USA}
\newcommand{\Amherst}{Department of Physics \& Astronomy, Amherst College, 25 East Drive, Amherst, MA 01003, USA}

\newcommand{\targetstar}{HD\,86728}
\newcommand{\planet}{HD\,86728\,b}

\shorttitle{NETS: HD 86728}
\shortauthors{Gupta et al}

\submitjournal{AJ}

\begin{document}

\title{The NEID Earth Twin Survey. I. Confirmation of a 31-day planet orbiting HD\,86728}

\correspondingauthor{Arvind F.\ Gupta}
\email{arvind.gupta@noirlab.edu}

\author[0000-0002-5463-9980]{Arvind F.\ Gupta}
\affil{\NOAO}

\author[0000-0002-4927-9925]{Jacob K. Luhn}
\affil{\UCI}

\author[0000-0001-6160-5888]{Jason T.\ Wright}
\affil{\PSUAA}
\affil{\PSUCEHW}
\affil{\PSETI}

\author[0000-0001-9596-7983]{Suvrath Mahadevan}
\affil{\PSUAA}
\affil{\PSUCEHW}

\author[0000-0003-0149-9678]{Paul Robertson}
\affil{\UCI}

\author[0000-0001-9626-0613]{Daniel M.\ Krolikowski}
\affil{\UA}

\author[0000-0001-6545-639X]{Eric B.\ Ford}
\affil{\PSUAA}
\affil{\PSUCEHW}
\affil{\PSUICS}
\affil{\PSUCASt}

\author[0000-0003-4835-0619]{Caleb I. Ca\~nas}
\affil{\GSFC}

\author[0000-0003-1312-9391]{Samuel Halverson}
\affil{\JPL}

\author[0000-0002-9082-6337]{Andrea S.J.\ Lin}
\affil{\PSUAA}
\affil{\PSUCEHW}

\author[0000-0001-8401-4300]{Shubham Kanodia}
\affil{\Carnegie}

\author[0000-0003-0199-9699]{Evan Fitzmaurice}
\affil{\PSUAA}
\affil{\PSUCEHW}
\affil{\PSUICDS}

\author[0000-0002-1743-3684]{Christian Gilbertson}
\affiliation{Center for Computing Research, Sandia National Laboratories, 1450 Innovation Pkwy SE, Albuquerque, NM 87123 USA}
\affil{\PSUAA}
\affil{\PSUCEHW}

\author[0000-0003-4384-7220]{Chad F.\ Bender}
\affil{\UA}

\author[0000-0002-6096-1749]{Cullen H.\ Blake}
\affil{\Penn}

\author[0000-0002-3610-6953]{Jiayin Dong}
\affil{\FlatironCCA}
\affil{\UIUC}

\author[0000-0002-0078-5288]{Mark R.~Giovinazzi}
\affil{\Penn}
\affil{\Amherst}

\author[0000-0002-9632-9382]{Sarah E.\ Logsdon}
\affil{\NOAO}


\author[0000-0002-0048-2586]{Andrew Monson}
\affil{\UA}

\author[0000-0001-8720-5612]{Joe P.\ Ninan}
\affil{\TIFR}

\author[0000-0002-2488-7123]{Jayadev Rajagopal}
\affil{\NOAO}

\author[0000-0001-8127-5775]{Arpita Roy}
\affiliation{Astrophysics \& Space Institute, Schmidt Sciences, New York, NY 10011, USA}

\author[0000-0002-4046-987X]{Christian Schwab}
\affil{\Macquarie}

\author[0000-0001-7409-5688]{Guðmundur Stefánsson}
\affil{\UAm}

\begin{abstract}

With close to three years of observations in hand, the NEID Earth Twin Survey (NETS) is starting to unearth new astrophysical signals for a curated sample of bright, radial velocity (RV)-quiet stars. We present the discovery of the first NETS exoplanet, HD\,86728\,b, a $m_p\sin i = 9.16^{+0.55}_{-0.56}\ \rm{M}_\oplus$ planet on a circular, $P=31.1503^{+0.0062}_{-0.0066}$ d orbit, thereby confirming a candidate signal identified by \citet{Hirsch2021}. We confirm the planetary origin of the detected signal, which has a semi-amplitude of just $K=1.91^{+0.11}_{-0.12}$ m~s$^{-1}$, via careful analysis of the NEID RVs and spectral activity indicators, and we constrain the mass and orbit via fits to NEID and archival RV measurements. The host star is intrinsically quiet at the $\sim1$ m~s$^{-1}$ level, with the majority of this variability likely stemming from short-timescale granulation. HD\,86728\,b is among the small fraction of exoplanets with similar masses and periods that have no known planetary siblings.

\end{abstract}


\section{Introduction} \label{sec:intro}

The newest generation of extreme precision radial velocity (EPRV) spectrographs has begun to deliver on its promise of achieving sub-m s$^{-1}$ on-sky precision \citep{Brewer2020,John2023} to enable the discovery and confirmation of low-mass exoplanets \citep{Barros2022}. 
As EPRV data sets accumulate through both uninformed exoplanet searches \citep{Motalebi2015,Hall2018,Hojjatpanah2019,Brewer2020}, follow-up observations of small transiting planets \citep{Demangeon2021,Brinkman2023}, and the continued development of cutting-edge RV analysis techniques \citep[e.g.,][]{Dumusque2018,Bedell2019,CollierCameron2021,deBeurs2024,Haywood2022,AlMoulla2024,Palumbo2024,Siegel2022}, we expect achieved RV sensitivity limits to continue to improve.

The NEID Earth Twin Survey \citep[NETS;][]{Gupta2021} is a dedicated search for low-mass exoplanets using the NEID EPRV spectrograph \citep{Schwab2016, Halverson2016}. 
Using time allocated through a guaranteed time observations (GTO) program, supplemented in part by Pennsylvania State University institutional access to NEID, we have been monitoring a carefully selected sample of 41 bright, RV-quiet stars since the commencement of this survey in September 2021. Observations from this survey have already been used to advance our understanding of stellar variability and spectral analysis techniques in the EPRV regime \citep[e.g.,][]{Burrows2024, Siegel2024}. Among the NETS targets is the bright ($V=5.4$ mag), nearby ($d = 14.9$ pc) star \targetstar\ (GJ\,376\,A, HIP\,49081). \citet{Gupta2021} ranked this star as the 5$^{\rm th}$ highest priority target for NETS based on the stellar properties, archival RV measurements, and empirical RV scatter \citep[RMS $= 3.2$ m~s$^{-1}$;][]{Butler2017}. With an Earth-equivalent insolation distance (EEID) of $r_{\rm EEID}=1.17$ au ($\theta_{\rm EEID}=78.4$ mas), \targetstar\ is also among the most promising targets for direct imaging searches for habitable-zone exoplanets. \citet{Mamajek2024} rank this star among the top 100 targets for the Habitable Worlds Observatory exo-Earth survey.

\targetstar\ has an extensive precision RV observing history with hundreds of measurements over more than three decades \citep{Butler2017,Rosenthal2021}.  Based on an analysis of RV data from the HIRES, Automated Planet Finder (APF), and Lick-Hamilton spectrographs, \citet{Hirsch2021} identified a planet candidate with a 31-day period and a RV semi-amplitude of $K = 2$ m~s$^{-1}$.

In this work, we present the first three years of NETS observations of the \targetstar\ system and we confirm the detection of the exoplanet \planet. We describe the new and existing RV observations in Section 2, the properties of the host star in Section 3, and our analysis of the full suite of data in Section 4. In Section 5, we discuss the properties of the newly discovered exoplanet and we place it in the context of the broader exoplanet population. We provide a summary of our results and conclusions in Section 6.

\section{Observations}\label{sec:obs}

\subsection{NEID}

NEID is a fiber-fed \citep{Kanodia2023}, environmentally-stabilized \citep{Robertson2019} optical-NIR echelle spectrograph installed on the WIYN\footnote{The WIYN Observatory is a joint facility of the NSF's National Optical-Infrared Astronomy Research Laboratory, Indiana University, the University of Wisconsin-Madison, Pennsylvania State University, Purdue University and Princeton University.} 3.5\,m telescope at Kitt Peak National Observatory. \targetstar\ was observed with NEID on 137 separate nights during the first three years of the NETS program. Each exposure was scheduled to trigger on a fixed signal-to-noise ratio (S/N) corresponding to a measurement precision of 30 cm s$^{-1}$, accounting for contributions from photon noise and the predicted residual oscillation signal as described by \citet{Gupta2021}. We collected one exposure per night except for the nights of 2021 February 11 and 28 and 2021 March 3 and 6, on which two visits with two back-to-back exposures were collected, and 2021 February 22, on which we obtained one pair of back-to-back exposures.

The 2D echelle spectra were processed with version $1.3$ of the NEID Data Reduction Pipeline\footnote{\url{https://neid.ipac.caltech.edu/docs/NEID-DRP/}} (DRP), producing wavelength-calibrated 1D spectra, RV measurements and uncertainties, and other higher level data products including stellar activity indicators. We discard one point flagged by the DRP due to an issue with the nightly wavelength solution stemming from a temporary failure of the laser frequency comb (LFC) on 2022 December 22, two more points on 2023 February 10 and 2023 March 28 for which we independently identified order-level RV outliers caused by LFC flux variations, and one outlier with S/N $< 100$ at 5500 \AA.
We are left with 42 nights (56 measurements) of high-quality data during the first year of the survey (hereafter Run 1; February 2021 - May 2022; median S/N $=336$) and 91 nights (91 measurements) of high-quality data during the following two years, after the instrument was thermally cycled following the Contreras fire at Kitt Peak (hereafter Run 2; December 2022 - May 2024; median S/N $=341$).

\begin{figure*} 
    \centering
    \includegraphics[trim={0 1.5cm 0 0},clip]{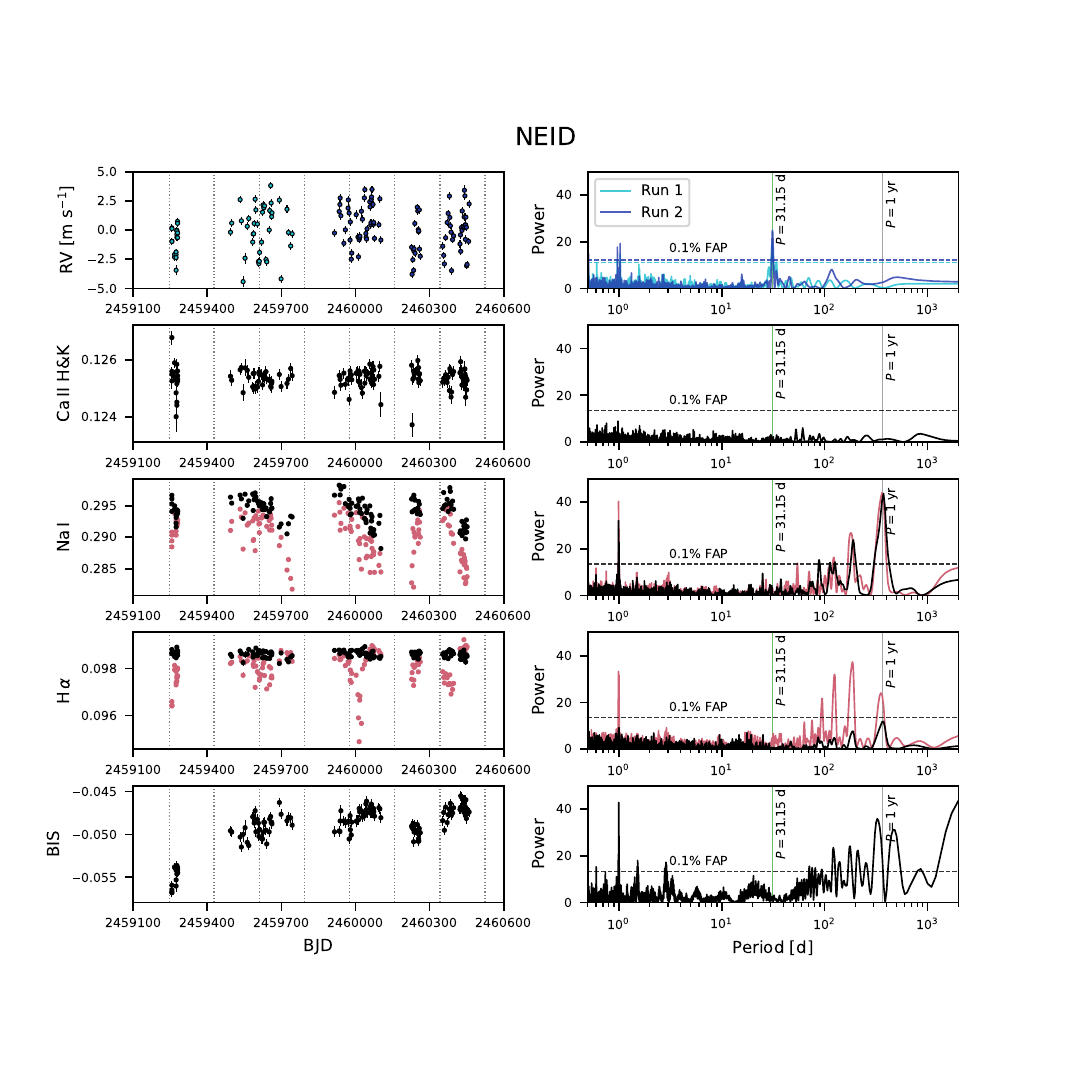}
    \caption{NEID RVs and activity indicator measurements for \targetstar\ (left). Vertical lines mark the boundaries between NEID's 6-month observing semesters. The GLS periodograms are shown to the right of each time series. We indicate the period of the detected RV signal with a green vertical line and Earth's orbital period with a grey vertical line, and the $0.1\%$ false alarm level with a horizontal dashed line. Periodograms are computed separately for the NEID RVs obtained during Run 1 (light blue) and Run 2 (dark blue). The pink points (left) and lines (right) show the pre-telluric correction Na I and H$\upalpha$ indicators, and the black points and lines show the post-correction indicators.}
    \label{fig:periodograms}
\end{figure*}

The NEID DRP calculates RVs using the cross-correlation function \citep[CCF;][]{Baranne1996} method, first computing separate CCFs for echelle orders 58--163 ($\sim380$\ nm -- $1.1\upmu$m) using curated line masks, and then fitting a Gaussian function to the peak of the weighted sum of the order-level CCFs to determine the overall RV of the spectrum. We found that due to an inconsistency in the construction of the reference wavelength solutions before and after the Contreras fire in 2022, a fixed zero point offset was introduced in the CCFs of the blue orders (echelle orders 119--163; $\sim380$--$520$\ nm). The wavelength solutions for these orders are anchored to the Thorium-Argon lamp rather than the LFC, which is used for the redder orders. We apply corrections to the order-level CCFs as described in the Appendix and then re-compute the weighted RVs following the standard pipeline algorithms. These velocities are hereafter referred to as the ``raw'' RVs. We obtain a median RV uncertainty of 33 cm s$^{-1}$ for the Run 1 data and 32 cm s$^{-1}$ for the Run 2 data. Although we have applied a first order correction, we still treat these separate runs as independent data sets for the remainder of this work, as major instrument interventions are observed to have an effect on the RVs of instruments with sub-m s$^{-1}$ precisions \cite[e.g.,][]{John2023}. The NEID RVs are shown in \autoref{fig:periodograms}.

We also note that version $1.3$ of the NEID DRP does not correct for telluric contamination when computing activity indicators. We therefore apply a correction  by first dividing the blaze-corrected spectra by each observation's telluric model, which includes both telluric line absorption and continuum absorption and is computed and provided as part of the DRP. We then rederive the activity indicators from the telluric-corrected spectra. The telluric model is only applicable to features in wavelength regions covered by the LFC spectrum (echelle orders 63--118; $\sim520$ nm--$1.1\ \upmu$m). The Ca II H \& K values are thus unaffected by this correction; the DRP values should be reliable as telluric contamination at these wavelengths is minimal. We show the pre- and post-correction H$\upalpha$ and Na I indicators in \autoref{fig:periodograms}. The telluric correction described here is set to be incorporated in a forthcoming version of the DRP.

\subsection{California Legacy Survey Data}

\begin{figure*} 
    \centering
    \includegraphics[trim={0 1.5cm 0 0},clip]{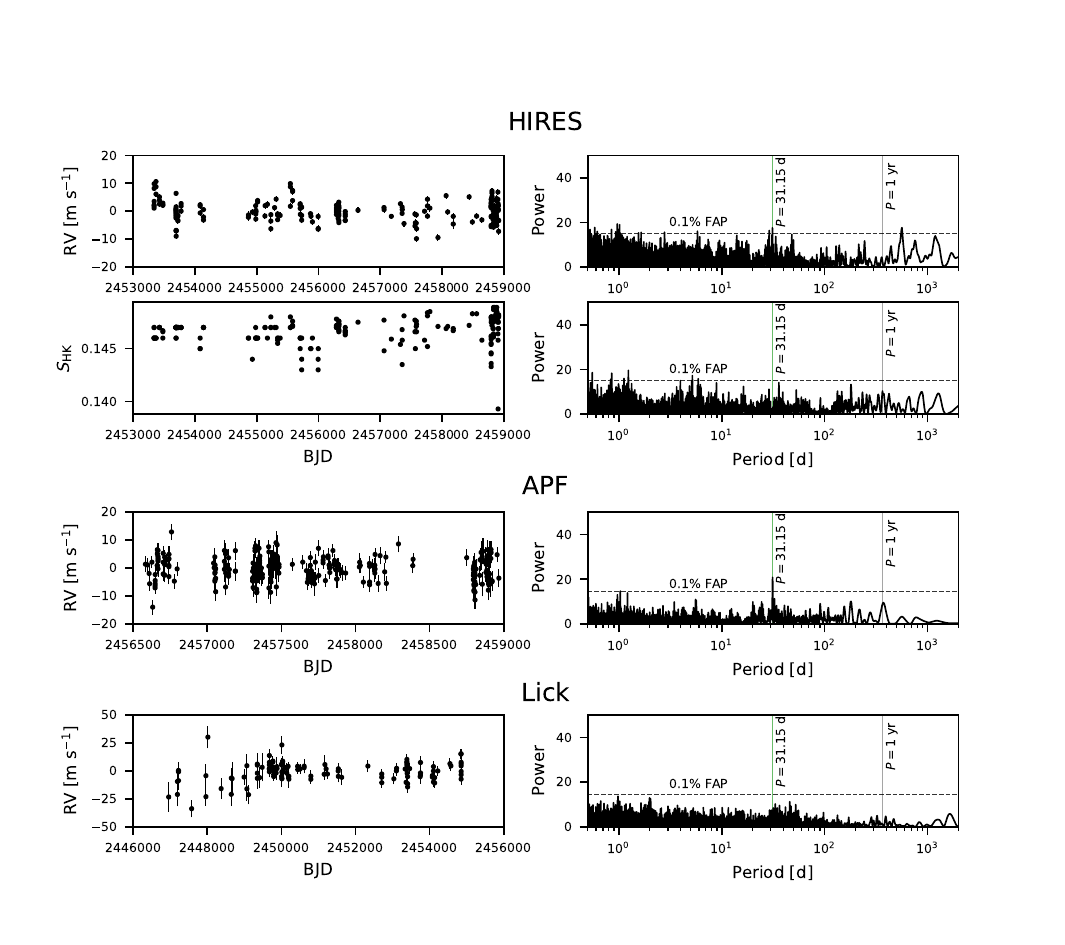}
    \caption{HIRES, APF, and Lick RVs and HIRES S$_{\rm HK}$ measurements for \targetstar. The GLS periodograms are shown to the right of each time series. As in \autoref{fig:periodograms}, we indicate the period of the detected RV signal with a green vertical line and Earth's orbital period with a grey vertical line, and the $0.1\%$ false alarm level with a horizontal dashed line.}
    \label{fig:cps_periodograms}
\end{figure*}

We also make use of RV measurements of \targetstar\ taken with the HIRES, APF, and Lick-Hamilton spectrographs and published as part of the California Legacy Survey (CLS) data set in \citet{Rosenthal2021}. For each of these instruments, RVs were calculated using the iodine technique \citep{Butler1996}, in which a molecular iodine spectrum projected onto the stellar spectrum is used as a reference for calculating Doppler velocity shifts. We show the RVs for these observations in \autoref{fig:cps_periodograms} and we analyze these alongside the NEID data in the following sections.

\subsubsection{Keck-HIRES}

The CLS data set includes measurements of \targetstar\ from the HIRES spectrograph \citep{Vogt1994} on the Keck I telescope taken between November 2004 and March 2020. These include 195 observations on 93 separate nights, with $120<$ S/N $<330$ and a median single measurement precision of $1.05$ m~s$^{-1}$. 

\subsubsection{Automated Planet Finder}

\targetstar\ was observed with the APF Levy spectrograph on the 2.4\,m robotic telescope at Mount Hamilton \citep{Vogt2014} as part of the APF-50 Survey \citep{Fulton2017}. A total of 289 individual RV measurements across 169 individual nights from October 2013 through April 2020 were published. These data have  $50<$ S/N $<200$ and a median single measurement precision of $2.82$ m~s$^{-1}$. 

\subsubsection{Lick-Hamilton}

\targetstar\ was also observed with the Hamilton spectrograph at Lick Observatory as part of the Lick Planet Search program \citep{Fischer2014}. The star was observed 117 times on 79 separate nights between February 1988 and January 2009, with $60<$ S/N $<380$ and a median single measurement precision of $5.16$ m~s$^{-1}$. 

\section{Stellar Properties}

We adopt the fundamental stellar parameters derived by \citet{Soubiran2024} for \targetstar. These parameters are listed in \autoref{tab:stellar_params}. \targetstar\ has an effective temperature of $T_{\rm eff} = 5610\pm46$ K and a surface gravity of $\log g = 4.25\pm0.02$, and it is likely evolved, with an age of $6-9$ Gyr \citep{Barry1988,Ng1998}. The star has a faint, widely separated M-dwarf binary companion, GJ\,376\,B, first identified using Two Micron All Sky Survey \citep[2MASS;][]{Skrutskie2006} data by \citet{Gizis2000}. But at a separation of $2005$ au (134''), GJ 376 B is not expected to induce a detectable RV signal for \targetstar.

\begin{deluxetable*}{lccc}
\tablecaption{Summary of Stellar Parameters for \targetstar \label{tab:stellar_params}}
\tablehead{\multicolumn{4}{l}{\hspace{-0.2cm} Identifiers: \targetstar, GJ\,376\,A, HIP\,49081, Gaia\,DR3\,746545172372256384} \\
\tableline
\colhead{~~~Parameter}&  \colhead{Value}&
\colhead{Description}&
\colhead{Reference}}
\startdata
\multicolumn{4}{l}{\hspace{-0.2cm} Coordinates and Parallax:} \\
~~~$\alpha_{\mathrm{J2016}}$ &    10:00:59.99 & Right Ascension (RA) & Gaia DR3\\
~~~$\delta_{\mathrm{J2016}}$ &   +31:55:18.34 & Declination (Dec) & Gaia DR3\\
~~~$\varpi$  & $66.9958 \pm 0.0092$ &  Parallax (mas) & Gaia DR3 \\
~~~$d$  & $14.92\pm0.02$ &  Distance (pc) & BJ21 \\
\multicolumn{4}{l}{\hspace{-0.2cm} Stellar Parameters:}\\
~~~$T_{\rm eff}$ & $5610\pm46$ & Effective Temperature (K) &  S24\\
~~~$L_\star$ & $1.365\pm0.014$ &  Luminosity (${\rm L}_\odot$) & S24\\
~~~$R_\star$ & $1.237\pm0.019$ & Stellar Radius (${\rm R}_\odot$)   &  S24\\
~~~$M_\star$ & $0.967\pm0.010$ & Stellar Mass (${\rm M}_\odot$)  &  S24\\
~~~$\log g$ & $4.25\pm0.02$ & Surface Gravity ($\log$ (cm~s$^{-2}$)) &  S24\\
~~~[Fe/H]& $0.20\pm0.02$ & Metallicity (dex) &  S24\\
~~~$v \sin i$ & $2.4\pm0.5$ & Rotational Broadening (km~s$^{-1}$) &  B16\\
\enddata
\tablenotetext{}{References: Gaia DR3 \citep{GaiaCollaboration2022}, S24 \citep{Soubiran2024}, B16 \citep{Brewer2016}, BJ21 \citep{Bailer-Jones2021}}
\end{deluxetable*}
\subsection{Stellar Rotation Period}\label{sec:rotation}

\citet{Hirsch2021} report a 32.6-day peak in the HIRES $S_{\rm HK}$ time series, which is a normalized measure of the Ca II H \& K line fluxes \citep{Baliunas1995}. Although they find no correlation between RV and $S_{\rm HK}$, the proximity of the activity signal to the 31-day RV signal raises concern that the RV variation is driven by stellar activity. 
We calculate the generalized Lomb-Scargle \citep[GLS;][]{Zechmeister2009,Zechmeister2018gls} periodogram of the Ca II H \& K, Na I, and H$\alpha$ activity indicators from the NEID spectroscopic time series for \targetstar\ to determine whether any rotational modulation can be detected in the stellar activity signal. These activity indicator measurements and periodograms are shown in \autoref{fig:periodograms}, with the \citet{Horne1986} power normalization. We also compute and show the time series and GLS for the bisector inverse slope (BIS) of the NEID CCFs. As in \citet{Burrows2024}, the BIS is calculated as the difference between the mean velocities of the 10${\rm th}$-40${\rm th}$ and 60${\rm th}$-90${\rm th}$ percentile depths of the CCF bisectors. While no signals are detected in the Ca II H \& K time series, the Na I, H$\upalpha$, and BIS periodograms exhibit power at Earth's orbital period of 1 year as well as at the $\frac{1}{2}$-year alias of this signal. Power is expected to show up at these frequencies due to the window function of the seasonal observing cadence, but we also note that these signals appear with different amplitudes in the pre- and post-telluric correction time series for Na I and H$\upalpha$. The peaks are more pronounced in the pre-correction data, which suggests that the signals are at least partly attributable to telluric contamination. The residual peaks in the post-correction data point to incomplete telluric removal. This is also consistent with the absence of an annual signal in Ca II H \& K, as tellurics are largely absent at these wavelengths. 
Given the lack of significant power at other periods in the telluric-corrected activity indicators or BIS, we do not detect the stellar rotation signal.

\targetstar\ was observed by the Transiting Exoplanet Survey Satellite \citep[TESS;][]{Ricker2015} in Sector 21 (2020 January 21 - 2020 February 18) during the primary mission and Sector 48 (2022 January 28 - 2022 February 26) during the extended mission. The combined differential photometric precisions in each of these sectors were 35 ppm and 34 ppm, respectively. We retrieve the processed TESS light curves for this target using \texttt{lightkurve} \citep{LightkurveCollaboration2018} and we use the Systematics-Insensitive Periodogram \citep[SIP;][]{Hedges2020} package to search for evidence of rotational modulation. The SIP analysis reveals no features in common between the two sectors and no significant periodicities above the noise level. The lack of a rotation signal in the photometry is not surprising; if the period were close to $30$ days as suggested by the \citet{Hirsch2021} S$_{\rm HK}$ detection, the TESS data would be insensitive to this signal due to the short baseline and standard pipeline detrending timescales.
If the period were instead much shorter than 30 days, the stellar spin axis must be highly inclined to avoid inconsistency with the small projected rotational velocity \citep[$v \sin i = 2.4$ km~s$^{-1}$,][]{Brewer2016}. We make no clear detection of the rotation period of \targetstar\ in the available time series data.

\section{Radial Velocity Analysis}

\subsection{Frequency Analysis}\label{sec:freq}

The GLS periodograms of the NEID and APF RVs are shown in \autoref{fig:periodograms}. In both data sets, we detect signals at $P\approx31$ d, consistent with the planet candidate reported by \citet{Hirsch2021}. These periodogram peaks have false alarm probabilities \citep[FAPs, calculated as in][]{Zechmeister2009} of $1.9\times10^{-12}$ (NEID Run 1), $5.7\times10^{-13}$ (NEID Run 2), and $9.4\times10^{-7}$ (APF), all of which are well below the conservative threshold of $0.1\%$ shown in \autoref{fig:periodograms}.

\begin{figure} 
    \includegraphics[trim={0.4cm 0.5cm 0 0},clip]{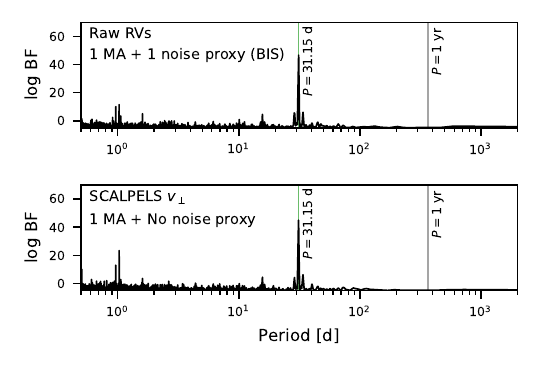} \caption{Bayes factor periodograms of NEID RVs following noise decorrelation with \texttt{Agatha}. The upper panel shows the results of the best-fit \texttt{Agatha} model for the raw RVs (single moving average term with BIS as the only noise proxy), and the lower panel shows the results of the best-fit \texttt{Agatha} model for the SCALPELS shift-driven RVs (single moving average term with no noise proxies).}
    \label{fig:agatha}
\end{figure}

Though the GLS periodogram is informative, one shortcoming of this model is that it is restricted to the assumption that the noise background is white, or uncorrelated in time. We assess the evidence for correlated noise in the NEID RVs using the \texttt{Agatha} framework \citep{Feng2017}, which accommodates a time-varying noise model. We compare a suite of models using different combinations of the BIS and Ca II H\&K, Na I, and H$\upalpha$ activity indicators as noise proxies, as well as models both with and without moving average components to describe correlations between consecutive data points. Based on the Bayes factor (BF), our \texttt{Agatha} analysis shows that the noise is best fit with a single moving average term the BIS as the lone noise proxy, though we note that this is only slightly better ($\delta \log BF = 1$) than the equivalent model with no noise proxies. The BF periodogram computed by \texttt{Agatha} for the best noise model is shown in \autoref{fig:agatha}. The 31-day peak has $\log BF = 46.7$.

\subsection{Keplerian-only fit}

\begin{figure*} 
\plottwo{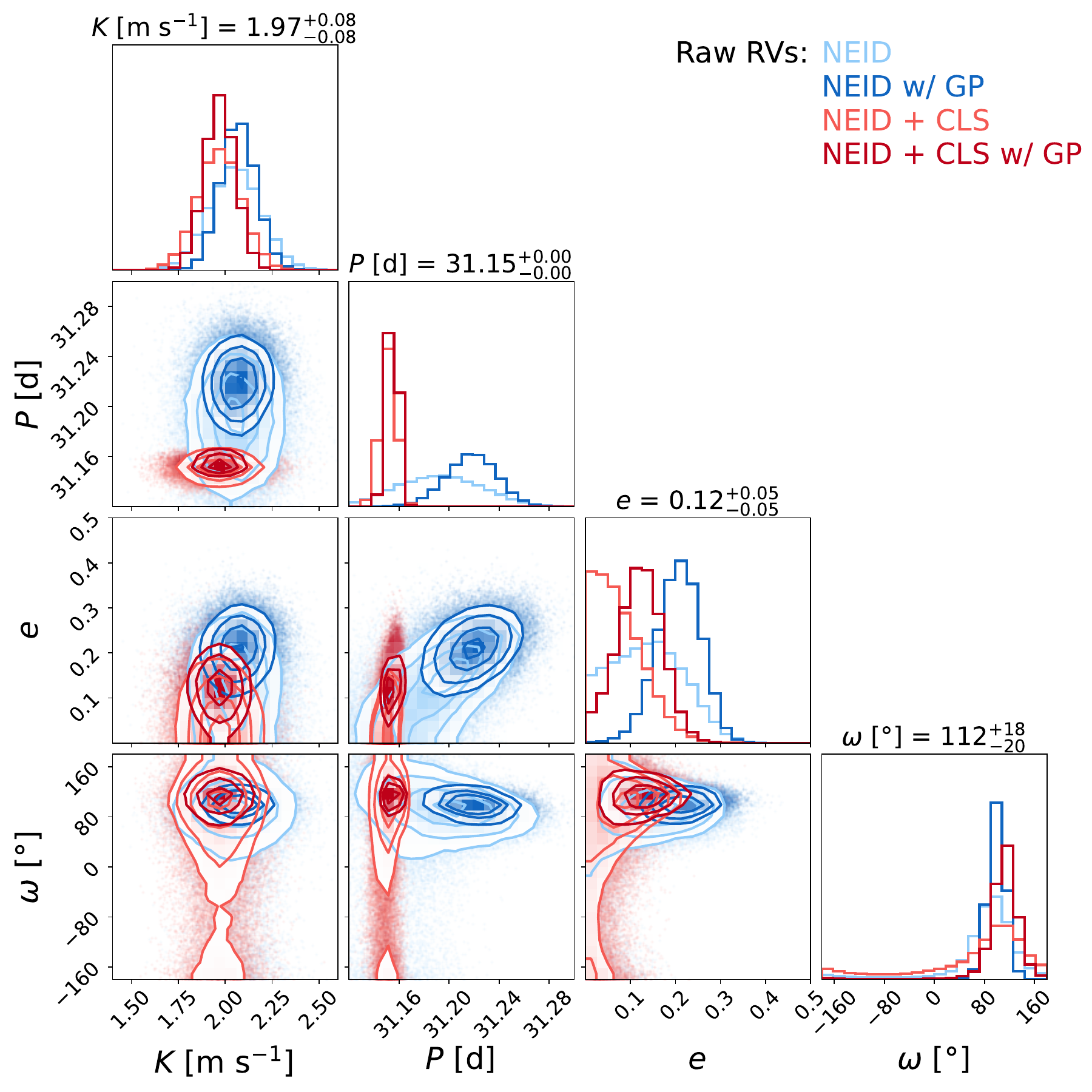}{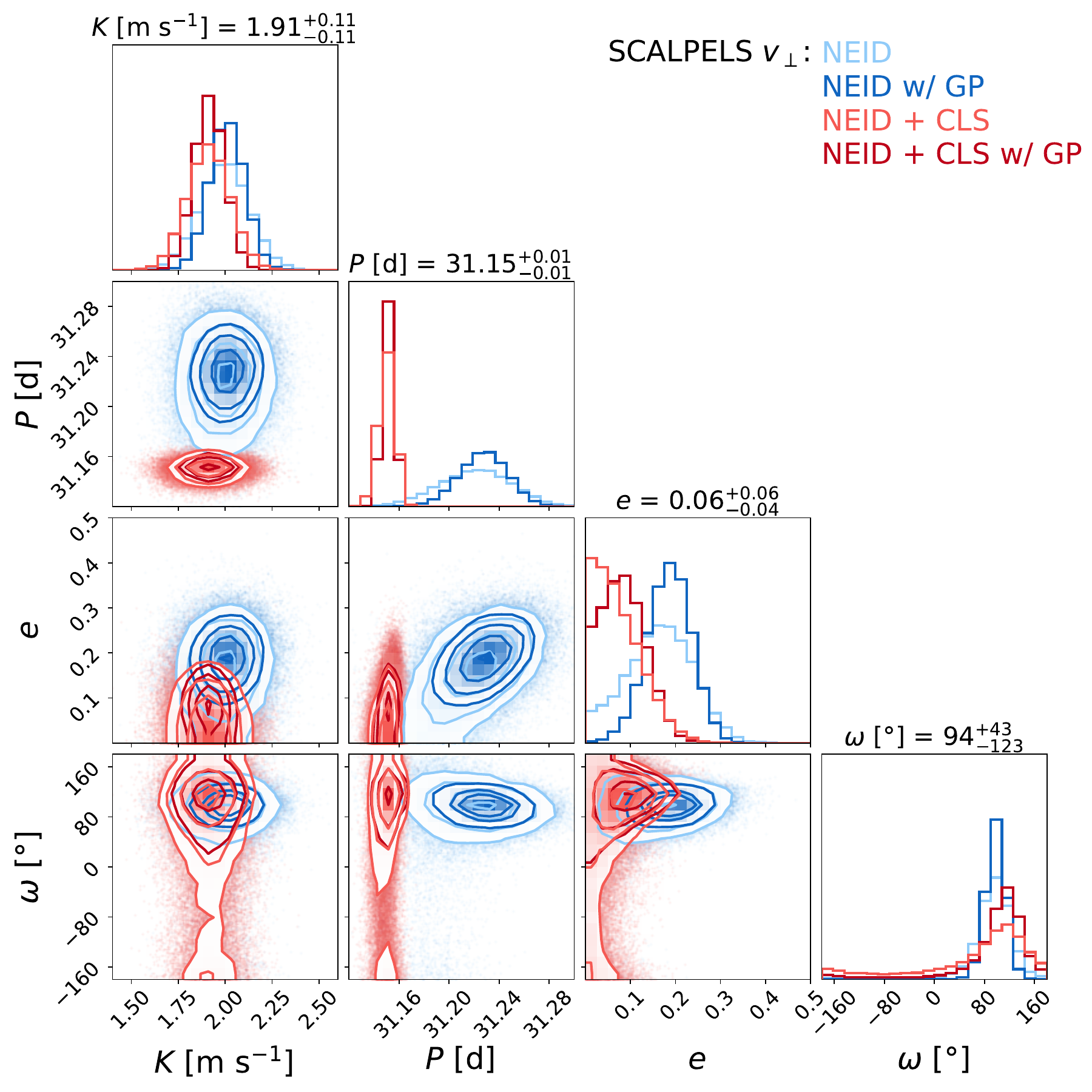}
\caption{Orbital parameter posteriors for \planet. We show the posteriors for each fit to the raw RVs (left) and to the SCALPELS $v_\perp$ (right). The 2D histograms include 1-, 2-, and 3-$\sigma$ contours. The best-fit parameter values and $\pm1$-$\sigma$ uncertainties are listed for our preferred models, which are the NEID + CLS fit with a GP for the raw NEID RVs, and the NEID + CLS fit without a GP to the SCALPELS $v_\perp$ time series.}
\label{fig:hd86728_corner}
\end{figure*}

We first fit a single Keplerian signal to the \targetstar\ NEID RV data using the \textsf{exoplanet} package \citep{Foreman-Mackey2021}, which uses the Hamiltonian Monte Carlo (HMC) package \texttt{PyMC3} \citep{Salvatier2016} for orbital parameter estimation. We model the orbit using a full Keplerian with a flat prior on the time of periastron ($t_p$), a log-uniform prior on semi-amplitude ($K$), and a broad uniform prior on the period. The eccentricity ($e$) and argument of periastron ($\omega$) are sampled on the unit disk in $\sqrt{e} \sin \omega -  \sqrt{e} \cos \omega$ space. Separate white noise jitter ($\sigma$) and systemic velocity ($\gamma$) terms are fit to each of the Run 1 and Run 2 time series. We assess sampling convergence by calculating the Gelman-Rubin statistic, $\hat{R}$, and requiring that $\hat{R}<1.01$ for all parameters in the fitting basis.
We show the best-fit orbital parameters in \autoref{tab:HD86728_fit_params_raw} and the posteriors in \autoref{fig:hd86728_corner}.

\begin{deluxetable*}{lrrrrrl}
\tablecaption{Derived parameters for \planet\ (raw NEID RVs)\label{tab:HD86728_fit_params_raw}}
\tablehead{\colhead{Parameter}& \colhead{Prior} & \multicolumn{4}{c}{Value} & 
\colhead{Unit}\\ & & NEID & NEID w/ GP & NEID + CLS & NEID + CLS w/ GP & }
\startdata
\multicolumn{7}{l}{\hspace{-0.2cm} Exoplanet Parameters:}  \\
~~~ $T_0$ & $\mathcal{U}(2459250,2459270)$ & $2459259.9^{+0.97}_{-0.99}$ & $2459259.45^{+0.62}_{-0.59}$ & $2459260.69^{+0.51}_{-0.49}$ & $2459260.89^{+0.37}_{-0.38}$ & BJD\\
~~~ $P$ & $\mathcal{U}(20,40)$ & $31.192^{+0.034}_{-0.034}$ & $31.218^{+0.02}_{-0.021}$ & $31.1515^{+0.0059}_{-0.0066}$ & $31.1548^{+0.0039}_{-0.0042}$ & days\\
~~~ $\sqrt{e}\cos\omega$ & $\mathcal{U}(-1.0,1.0)$ & $-0.02^{+0.17}_{-0.16}$ & $-0.08^{+0.11}_{-0.1}$ & $-0.05^{+0.17}_{-0.15}$ & $-0.13^{+0.11}_{-0.09}$ & \\
~~~ $\sqrt{e}\sin\omega$ & $\mathcal{U}(-1.0,1.0)$ & $0.33^{+0.11}_{-0.21}$ & $0.439^{+0.051}_{-0.062}$ & $0.18^{+0.14}_{-0.2}$ & $0.307^{+0.072}_{-0.1}$ & \\
~~~ $e$ & derived & $0.137^{+0.084}_{-0.086}$ & $0.209^{+0.047}_{-0.05}$ & $0.067^{+0.065}_{-0.046}$ & $0.121^{+0.048}_{-0.052}$ & \\
~~~ $\omega$ & derived & $91^{+30}_{-38}$ & $100^{+13}_{-14}$ & $92^{+43}_{-113}$ & $112^{+18}_{-20}$ & $^\circ$\\
~~~ $\log K$ & $\mathcal{U}(-2,2)$ & $0.719^{+0.065}_{-0.07}$ & $0.727^{+0.044}_{-0.047}$ & $0.677^{+0.057}_{-0.061}$ & $0.678^{+0.041}_{-0.042}$ & log m~s$^{-1}$\\
~~~ $K$ & derived & $2.05^{+0.14}_{-0.14}$ & $2.069^{+0.093}_{-0.094}$ & $1.97^{+0.12}_{-0.12}$ & $1.971^{+0.082}_{-0.081}$ & m~s$^{-1}$\\
~~~ $a$ & derived & $0.19177^{+0.00068}_{-0.00068}$ & $0.19187^{+0.00066}_{-0.00067}$ & $0.1916^{+0.00066}_{-0.00065}$ & $0.19161^{+0.00066}_{-0.00067}$ & au\\
~~~ $m_p \sin i$ & derived & $9.76^{+0.67}_{-0.66}$ & $9.73^{+0.45}_{-0.45}$ & $9.43^{+0.56}_{-0.57}$ & $9.4^{+0.4}_{-0.4}$ & M$_\oplus$\\
\multicolumn{7}{l}{\hspace{-0.2cm} Instrument and Model Parameters:}  \\
~~~ $\gamma_{\rm NEID, Run1}$  & $\mathcal{N}(0,100)$ & $-0.41^{+0.14}_{-0.14}$ & $-0.46^{+0.12}_{-0.12}$ & $-0.37^{+0.14}_{-0.14}$ & $-0.37^{+0.13}_{-0.13}$ & m~s$^{-1}$\\
~~~ $\gamma_{\rm NEID, Run2}$  & $\mathcal{N}(0,100)$ & $0.0^{+0.13}_{-0.13}$ & $-0.02^{+0.11}_{-0.11}$ & $0.04^{+0.13}_{-0.13}$ & $0.03^{+0.12}_{-0.12}$ & m~s$^{-1}$\\
~~~ $\gamma_{\rm HIRES}$  & $\mathcal{N}(0,100)$ & $$ & $$ & $-0.07^{+0.25}_{-0.25}$ & $-0.03^{+0.19}_{-0.19}$ & m~s$^{-1}$\\
~~~ $\gamma_{\rm APF}$  & $\mathcal{N}(0,100)$ & $$ & $$ & $-0.29^{+0.21}_{-0.21}$ & $-0.28^{+0.17}_{-0.18}$ & m~s$^{-1}$\\
~~~ $\gamma_{\rm Lick}$  & $\mathcal{N}(0,100)$ & $$ & $$ & $0.88^{+0.7}_{-0.71}$ & $0.71^{+0.5}_{-0.51}$ & m~s$^{-1}$\\
~~~ $\sigma_{\rm NEID, Run1}$ & $\mathcal{U}(0,25)$ & $0.96^{+0.12}_{-0.1}$ & $0.883^{+0.077}_{-0.07}$ & $0.98^{+0.12}_{-0.1}$ & $0.916^{+0.08}_{-0.073}$ & m~s$^{-1}$\\
~~~ $\sigma_{\rm NEID, Run2}$ & $\mathcal{U}(0,25)$ & $1.15^{+0.1}_{-0.09}$ & $1.076^{+0.068}_{-0.063}$ & $1.15^{+0.1}_{-0.09}$ & $1.085^{+0.069}_{-0.064}$ & m~s$^{-1}$\\
~~~ $\sigma_{\rm HIRES}$  & $\mathcal{U}(0,25)$ & $$ & $$ & $3.31^{+0.19}_{-0.18}$ & $3.2^{+0.14}_{-0.13}$ & m~s$^{-1}$\\
~~~ $\sigma_{\rm APF}$  & $\mathcal{U}(0,25)$ & $$ & $$ & $2.28^{+0.24}_{-0.24}$ & $2.19^{+0.17}_{-0.17}$ & m~s$^{-1}$\\
~~~ $\sigma_{\rm Lick}$  & $\mathcal{U}(0,25)$ & $$ & $$ & $4.95^{+0.81}_{-0.77}$ & $4.74^{+0.58}_{-0.57}$ & m~s$^{-1}$\\
~~~ $\log\sigma_{\rm GP}$  & $\mathcal{U}(-5,5)$ & & $-0.76^{+0.31}_{-0.32}$ & & $-0.14^{+0.27}_{-0.34}$ & log m~s$^{-1}$\\
~~~ $\log\rho_{\rm GP}$  & $\mathcal{N}(4.0,1.3)$ & & $4.71^{+0.41}_{-0.41}$ & & $4.59^{+0.38}_{-0.37}$ & log days\\
\multicolumn{7}{l}{\hspace{-0.2cm}Detection Significance:}\\
~~~ $\Delta{\rm BIC}_{0-1}$ & & 191.4 & 210.6 & 202.3 & 233.4 & \\
\enddata
\end{deluxetable*}

\begin{figure*} 
    \includegraphics[width=\linewidth]{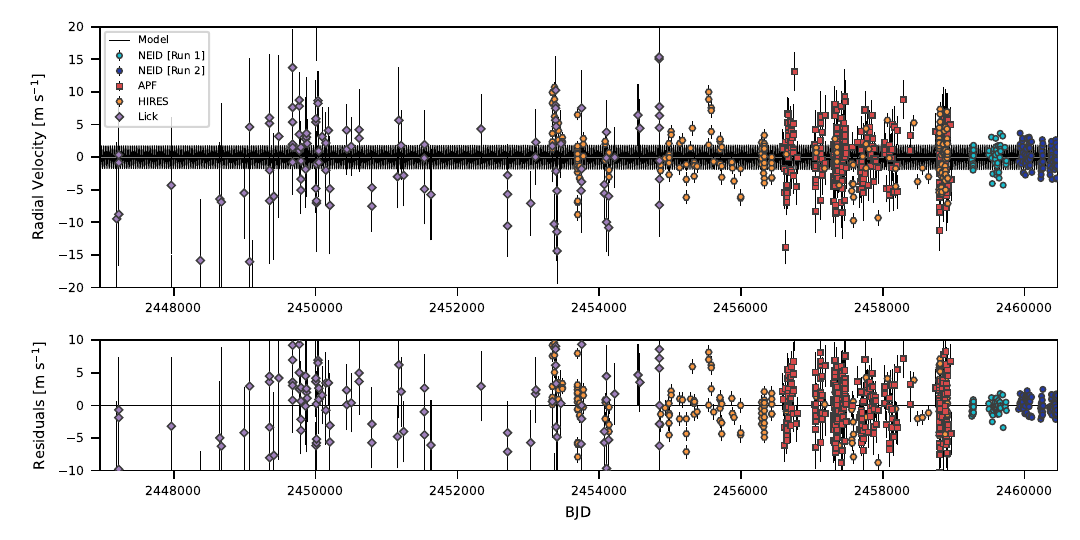} \caption{Complete RV time series for \targetstar. The NEID data are shown as light (Run 1) and dark (Run 2) blue circles, the APF data are shown as red squares, HIRES as orange hexagons, and Lick as purple diamonds. We also show the best-fit model (black), and in the lower panel we show the residuals to this model.}
    \label{fig:all_rvs}
\end{figure*}

We repeat the \textsf{exoplanet} orbit fit using the combined NEID + CLS data set (\autoref{fig:all_rvs}). We include additional jitter and offset terms for each of the CLS instruments (Lick, HIRES, and APF), but the fitting basis and priors are otherwise unchanged. In \autoref{fig:hd86728_corner} and \autoref{tab:HD86728_fit_params_raw}, we show that this joint fit is consistent with the NEID-only fit to better than 1-$\sigma$. The extended CLS observing baseline significantly tightens the precision on the orbital period and drags the eccentricity closer to zero, but the semi-amplitude constraint appears to be driven primarily by the higher precision NEID RV measurements.

\subsection{Keplerian+GP fit}

Given the absence of any periodic signals in the NEID activity indicators (Section \ref{sec:rotation}), the observed RVs are unlikely to be driven by rotationally-modulated variations. However, we note that the scatter in the NEID residuals to the best-fit planet signal is on the order of 1 m~s$^{-1}$, which greatly exceeds the measurement noise ($\sim$$0.3$ m~s$^{-1}$). This difference is reflected in the best-fit jitter terms for the NEID RVs, which are $\sigma_{\rm NEID,Run 1} = 0.96$ m~s$^{-1}$ and $\sigma_{\rm NEID,Run 2} = 1.15$ m~s$^{-1}$ (NEID-only fit) and $\sigma_{\rm NEID,Run 1} = 0.98$ m~s$^{-1}$ and $\sigma_{\rm NEID,Run 2} = 1.15$ m~s$^{-1}$ (NEID + CLS fit). Intrinsic stellar variability is still likely contributing to the observed RV scatter at the sub-m~s$^{-1}$ level even though we detect neither significant activity-RV correlations nor periodic activity signals for this star.

Here, we use Gaussian process (GP) regression to assess our sensitivity to correlated noise driven by stellar magnetic activity. We use a Mat\'ern-3/2 kernel:
\begin{equation}
    k_{M\,3/2}(\tau) = \sigma^2 \left(1 + \frac{\sqrt{3}\tau}{\rho}\right)e^{-\sqrt{3}\tau/\rho}
\end{equation}
with length scale hyperparameter $\rho$ and amplitude hyperparameter $\sigma$ to model correlations between observations separated by time $\tau$. This kernel was chosen for ease of implementation in the \textsf{exoplanet} fitting framework using \textsf{celerite2} \citep{Foreman-Mackey2017,Foreman-Mackey2018}. We condition the hyperparameters on the NEID BIS time series using \textsf{exoplanet} and PyMC3. The BIS is selected as a reasonable proxy for the activity level based on the \texttt{Agatha} analysis in Section \ref{sec:freq}. The hyperparameters are both sampled on log-normal distributions with broad priors. We then fit the GP with a Keplerian signal to the NEID RV time series, using the $\log \rho$ posteriors from the BIS fit as priors, and this is repeated for the full NEID + CLS RV time series as above. The results of these two fits with a GP are given in \autoref{tab:HD86728_fit_params_raw} and \autoref{fig:hd86728_corner}. They are largely consistent with each other, but we note that there is a $\sim3$-$\sigma$ discrepancy in the orbital period. When applied only to the NEID data, the GP drags the orbital period away from the more precise value we get when including the CLS data in the fit.  We compare the Keplerian-only fits to the Keplerian + GP fits in Section \ref{sec:modelcomp}.

\subsection{SCALPELS RV Analysis}

As an independent test of the contribution of intrinsic stellar variability to the observed RV signal, we analyze the autocorrelation function (ACF) of the NEID CCFs using the SCALPELS method described in \citet{CollierCameron2021}. The ACF is expected to be sensitive to asymmetries in the line profiles, which will probe RV perturbations resulting from stellar variability rather than bulk Doppler shifts. After applying the CCF shifts described in the Appendix, we calculate the ACFs of the weighted sums in the velocity range $\pm20$ km~s$^{-1}$, which is the same range used for the NEID DRP RV calculation. We then compute the singular value decomposition of the ACFs to extract the principal component basis vectors. Projecting the RVs onto these basis vectors gives the ``shape-driven'' component of the time series ($v_\parallel$), and projecting them onto the corresponding orthogonal subspace gives the ``shift-driven'' RV time series ($v_\perp$). We retain the 8 highest variance basis vectors, beyond which we see only marginal improvement in the reconstruction of the ACF. This approach allows us to isolate and remove the stellar variability contribution to the RVs, $v_\parallel$, from the bulk Doppler signal, $v_\perp$. We note that SCALPELS was applied only to the NEID data, as it is compatible only with CCF-based RVs and not with the iodine reference technique that was used for the other spectrographs.

\begin{figure*} 
    \includegraphics[width=\linewidth]{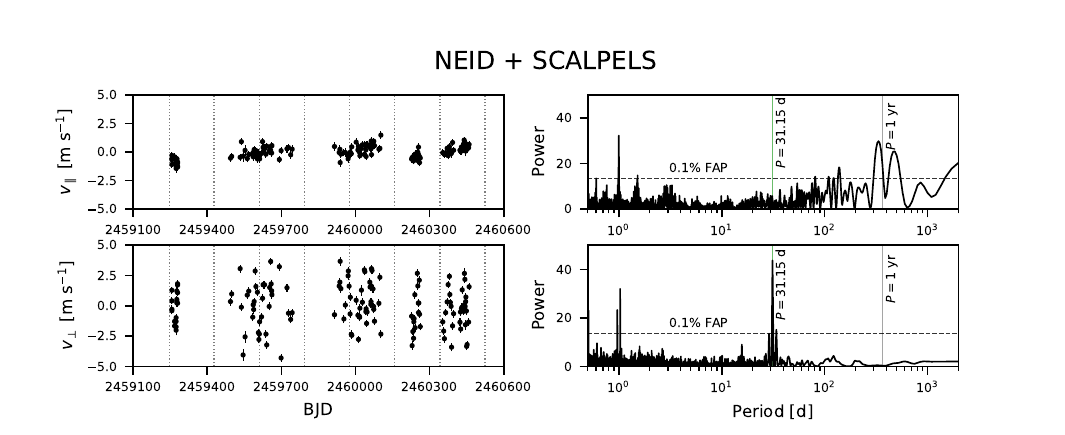} \caption{SCALPELS decomposition of the NEID RVs. We show the shape-driven velocity component, $v_\parallel$, and the corresponding GLS periodogram in the upper left and right, respectively. The lower panels show the shift-driven velocity component, $v_\perp$, and GLS periodogram. The 31-day RV signal is isolated almost entirely to the shift-driven component. The shape-driven component is dominated by a ~1 yr signal that closely resembles that seen in the BIS. This signal is not seen in the raw RVs.}
    \label{fig:scalpels}
\end{figure*}

The decomposed NEID velocities are shown in \autoref{fig:scalpels}. GLS periodogram analyses of these time series show that the 31-day signal seen in the raw RVs is shift-driven, not shape-driven, confirming that stellar variability is unlikely to be responsible. We repeat the \texttt{Agatha} analysis (Section \ref{sec:freq}) on the cleaned SCALPELS RVs, and we find that the best-fit noise model includes a single moving average term and no noise proxies. This suggests that the SCALPELS decomposition removed the RV variability being traced by the BIS, which is consistent with the similar frequency structure seen for the BIS and the SCALPELS shape-driven velocities.

All of the \textsf{exoplanet} fits described above (NEID-only and NEID + CLS, both with and without a GP) were repeated using the cleaned, shift-driven NEID time series from SCALPELS. Though the \texttt{Agatha} analysis suggests that a GP may not be necessary for the NEID RVs, we note that the shape-driven changes traced by the BIS may still be present in the CLS data, which were not cleaned with SCALPELS. The posteriors are shown in \autoref{fig:hd86728_corner} and listed in \autoref{tab:HD86728_fit_params_scalpels}. We find that the results are very similar, regardless of whether the analysis was performed using the raw NEID RVs or shift-driven RVs from SCALPELS.

\begin{deluxetable*}{lrrrrrl}
\tablecaption{Derived parameters for \planet\ (SCALPELS $v_\perp$)\label{tab:HD86728_fit_params_scalpels}}
\tablehead{\colhead{Parameter}& \colhead{Prior} & \multicolumn{4}{c}{Value} & 
\colhead{Unit}\\ & & NEID & NEID w/ GP & \textbf{NEID + CLS}$^\dagger$ & NEID + CLS w/ GP & }
\startdata
\multicolumn{7}{l}{\hspace{-0.2cm} Exoplanet Parameters:}  \\
~~~ $T_0$ & $\mathcal{U}(2459250,2459270)$ & $2459259.09^{+0.83}_{-0.79}$ & $2459258.99^{+0.57}_{-0.55}$ & \textbf{2459260.43}$^{+0.53}_{-0.48}$ & $2459260.53^{+0.42}_{-0.39}$ & BJD\\
~~~ $P$ & $\mathcal{U}(20,40)$ & $31.221^{+0.028}_{-0.03}$ & $31.228^{+0.019}_{-0.02}$ & \textbf{31.1503}$^{+0.0062}_{-0.0066}$ & $31.1516^{+0.0044}_{-0.0046}$ & days\\
~~~ $\sqrt{e}\cos\omega$ & $\mathcal{U}(-1.0,1.0)$ & $-0.05^{+0.16}_{-0.16}$ & $-0.06^{+0.11}_{-0.11}$ & \textbf{-0.06}$^{+0.17}_{-0.15}$ & $-0.1^{+0.15}_{-0.11}$ & \\
~~~ $\sqrt{e}\sin\omega$ & $\mathcal{U}(-1.0,1.0)$ & $0.377^{+0.087}_{-0.141}$ & $0.414^{+0.054}_{-0.067}$ & \textbf{0.17}$^{+0.14}_{-0.2}$ & $0.238^{+0.089}_{-0.145}$ & \\
~~~ $e$ & derived & $0.167^{+0.072}_{-0.081}$ & $0.187^{+0.047}_{-0.05}$ & \textbf{0.064}$^{+0.065}_{-0.044}$ & $0.08^{+0.053}_{-0.05}$ & \\
~~~ $\omega$ & derived & $96^{+24}_{-28}$ & $98^{+15}_{-16}$ & \textbf{94}$^{+43}_{-123}$ & $110^{+26}_{-44}$ & $^\circ$\\
~~~ $\log K$ & $\mathcal{U}(-2,2)$ & $0.692^{+0.064}_{-0.069}$ & $0.698^{+0.046}_{-0.047}$ & \textbf{0.648}$^{+0.058}_{-0.062}$ & $0.65^{+0.041}_{-0.043}$ & log m~s$^{-1}$\\
~~~ $K$ & derived & $2.0^{+0.13}_{-0.13}$ & $2.009^{+0.094}_{-0.093}$ & \textbf{1.91}$^{+0.11}_{-0.12}$ & $1.915^{+0.081}_{-0.081}$ & m~s$^{-1}$\\
~~~ $a$ & derived & $0.19188^{+0.00068}_{-0.00068}$ & $0.19191^{+0.00066}_{-0.00066}$ & \textbf{0.1916}$^{+0.00065}_{-0.00066}$ & $0.1916^{+0.00066}_{-0.00066}$ & au\\
~~~ $m_p \sin i$ & derived & $9.46^{+0.63}_{-0.64}$ & $9.49^{+0.45}_{-0.44}$ & \textbf{9.16}$^{+0.55}_{-0.56}$ & $9.17^{+0.4}_{-0.4}$ & M$_\oplus$\\
\multicolumn{7}{l}{\hspace{-0.2cm} Instrument and Model Parameters:}  \\
~~~ $\gamma_{\rm NEID, Run1}$  & $\mathcal{N}(0,100)$ & $-0.35^{+0.13}_{-0.13}$ & $-0.367^{+0.094}_{-0.093}$ & \textbf{-0.29}$^{+0.13}_{-0.13}$ & $-0.29^{+0.1}_{-0.1}$ & m~s$^{-1}$\\
~~~ $\gamma_{\rm NEID, Run2}$  & $\mathcal{N}(0,100)$ & $0.05^{+0.13}_{-0.13}$ & $0.05^{+0.093}_{-0.094}$ & \textbf{0.11}$^{+0.12}_{-0.12}$ & $0.107^{+0.099}_{-0.1}$ & m~s$^{-1}$\\
~~~ $\gamma_{\rm HIRES}$  & $\mathcal{N}(0,100)$ & $$ & $$ & \textbf{-0.06}$^{+0.25}_{-0.25}$ & $-0.05^{+0.18}_{-0.18}$ & m~s$^{-1}$\\
~~~ $\gamma_{\rm APF}$  & $\mathcal{N}(0,100)$ & $$ & $$ & \textbf{-0.29}$^{+0.21}_{-0.21}$ & $-0.29^{+0.16}_{-0.16}$ & m~s$^{-1}$\\
~~~ $\gamma_{\rm Lick}$  & $\mathcal{N}(0,100)$ & $$ & $$ & \textbf{0.86}$^{+0.69}_{-0.7}$ & $0.83^{+0.49}_{-0.49}$ & m~s$^{-1}$\\
~~~ $\sigma_{\rm NEID, Run1}$ & $\mathcal{U}(0,25)$ & $0.88^{+0.11}_{-0.1}$ & $0.849^{+0.074}_{-0.067}$ & \textbf{0.92}$^{+0.11}_{-0.1}$ & $0.902^{+0.077}_{-0.07}$ & m~s$^{-1}$\\
~~~ $\sigma_{\rm NEID, Run2}$ & $\mathcal{U}(0,25)$ & $1.109^{+0.099}_{-0.088}$ & $1.086^{+0.067}_{-0.063}$ & \textbf{1.12}$^{+0.1}_{-0.09}$ & $1.096^{+0.07}_{-0.064}$ & m~s$^{-1}$\\
~~~ $\sigma_{\rm HIRES}$  & $\mathcal{U}(0,25)$ & $$ & $$ & \textbf{3.3}$^{+0.2}_{-0.18}$ & $3.26^{+0.14}_{-0.14}$ & m~s$^{-1}$\\
~~~ $\sigma_{\rm APF}$  & $\mathcal{U}(0,25)$ & $$ & $$ & \textbf{2.27}$^{+0.24}_{-0.24}$ & $2.24^{+0.17}_{-0.17}$ & m~s$^{-1}$\\
~~~ $\sigma_{\rm Lick}$  & $\mathcal{U}(0,25)$ & $$ & $$ & \textbf{4.94}$^{+0.81}_{-0.78}$ & $4.85^{+0.57}_{-0.56}$ & m~s$^{-1}$\\
~~~ $\log\sigma_{\rm GP}$  & $\mathcal{U}(-5,5)$ &  & $-2.8^{+1.2}_{-1.5}$ &  & $-1.8^{+1.4}_{-2.1}$ & log m~s$^{-1}$\\
~~~ $\log\rho_{\rm GP}$  & $\mathcal{N}(4.0,1.3)$ &  & $4.65^{+0.43}_{-0.43}$ & & $4.6^{+0.43}_{-0.42}$ & log days\\
\multicolumn{7}{l}{\hspace{-0.2cm}Detection Significance:}\\
~~~ $\Delta{\rm BIC}_{0-1}$ & & 221.6 & 212.3 & \textbf{220.8} & 211.4 & \\
\enddata
\tablenotetext{}{$^\dagger$Adopted best-fit parameters for \planet}
\end{deluxetable*}

\subsection{Model Comparison}\label{sec:modelcomp}

We assess the results of each of our fits using Bayesian Information Criterion (BIC), which is calculated as
\begin{equation}
    {\rm BIC} = k \ln n_{\rm obs} - 2 \ln \mathcal{L}
\end{equation}
where $n_{\rm obs}$ is the number of observations used in the fit, $k$ is number of model parameters, and
$\mathcal{L}$ is the model likelihood. Fits with lower BIC values are preferred. Because this metric penalizes models with more parameters, which can achieve a better likelihood by overfitting, it provides a mechanism to compare models of different complexity.

We first compare the BIC of each fit to that of a zero-planet, constant velocity model. These values are listed at the bottom of Tables \ref{tab:HD86728_fit_params_raw} and \ref{tab:HD86728_fit_params_scalpels}. In all cases, the planet signal is strongly favored, with $\Delta{\rm BIC}_{0-1}>190$.
We then calculate the difference between the fits with and without a GP, focusing first on the fits to the raw RV time series. We find $\Delta{\rm BIC}_{\rm NEID} = 19.1$ and $\Delta{\rm BIC}_{\rm NEID+CLS} = 31.1$, with the metric favoring the Keplerian + GP fit in both cases. However, using the SCALPELS $v_\perp$ time series, we calculate $\Delta{\rm BIC}_{\rm NEID} = -9.4$ and $\Delta{\rm BIC}_{\rm NEID+CLS} = -9.3$, favoring the Keplerian-only model. That is, the addition of a GP does not appreciably improve the likelihood of each SCALPELS fit, so the added model parameters are not justified. The difference in model preference between the SCALPELS and raw RV fits is likely because much of the RV structure being captured by the GP in the raw fit is absent from the SCALPELS shift-driven RVs. This signal is being removed by the SCALPELS decomposition and assigned to the shape-driven time series.

\begin{figure} 
    \centering
    \includegraphics[width=\linewidth]{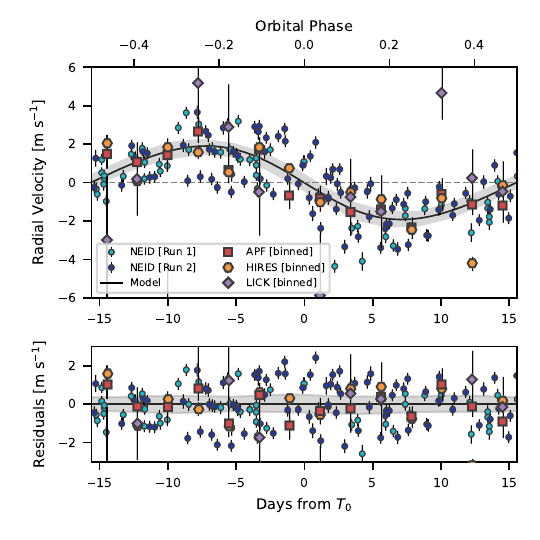}
    \caption{Phase-folded RVs for \targetstar\ (top) and residuals to the best-fit orbital solution for the combined data set (bottom). The best-fit model and 3-$\sigma$ uncertainties are shown in black and grey, respectively. We bin the CLS RVs by orbital phase to show that they are consistent with the fit. The NEID RVs and model fit shown here are based on the SCALPELS $v_\perp$ measurements.}
    \label{fig:hd86728_folded}
\end{figure}

The best-fit model parameters for all of the above analyses (with and without a GP; NEID-only and NEID+CLS; raw NEID RVs and SCALPELS $v_\perp$) are given in Tables \ref{tab:HD86728_fit_params_raw} and \ref{tab:HD86728_fit_params_scalpels}. With the exception of a 3-$\sigma$ discrepancy in orbital period between the NEID-only and NEID+CLS fits, the posteriors are consistent to better than 1-$\sigma$ across all fits. Taking the SCALPELS decomposition to have reliably isolated the exoplanet signal from most of the correlated noise, we adopt the results of the NEID+CLS, Keplerian-only fit with the SCALPELS shift-driven RVs as our preferred model for the remainder of this work. We find that \planet\ is a $m_p\sin i = 9.16^{+0.55}_{-0.56}\ \rm{M}_\oplus$ planet with a $P=31.1503^{+0.0062}_{-0.0066}$ d, $e=0.064^{+0.065}_{-0.044}$ orbit. We show the phase-folded best-fit orbit and the RVs and residuals in \autoref{fig:hd86728_folded}.

\subsection{Stellar Jitter}

\begin{figure} 
    \centering
    \includegraphics[width=\linewidth]{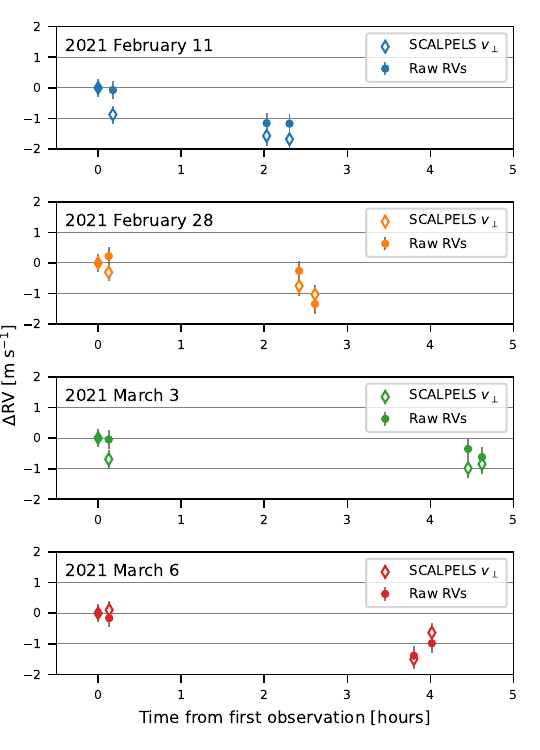}
    \caption{NEID RVs for the four nights with multiple visits: 2021 February 11, 2021 February 28, 2021 March 3, and 2021 March 6. We show the raw RVs as closed circles and the SCALPELS $v_\perp$ values as open diamonds. Velocities are plotted relative to the first measurement taken on each night. On each of these nights, the RV signal changes by $\sim1$ m~s$^{-1}$ between visits separated by multiple hours.}
    \label{fig:high_cadence}
\end{figure}

An important caveat to the SCALPELS analysis is that it is sensitive only to shape-driven variations that are detectable at the S/N and resolution of the input data. Because not all types of stellar variability will induce detectable line profile changes, the SCALPELS $v_\perp$ time series will retain some amount of stellar jitter. One plausible source of the excess RV scatter in the residuals is granulation, which operates on timescales ranging from several hours up to 1-2 days, and which has recently been shown to dominate the solar RV signature over inactive regions of the surface \citep{Lakeland2024}. The NEID RVs for the four nights with multiple visits are shown in \autoref{fig:high_cadence}. We find that on each of these nights, the RV signal does indeed change by about 1 m~s$^{-1}$ for measurements separated by 2-5 hours, pointing to short-timescale variability, such as granulation, as the main source of the observed jitter. This explanation is also consistent with the non-detection of stellar signals in the frequency domain; aside from these few nights, our observing cadence is too sparse to adequately sample the granulation signature, so it is largely indistinguishable from white noise throughout the remainder of the time series. We also show in \autoref{fig:high_cadence} that these intra-night RV changes are not removed by SCALPELS. This further supports granulation as the source of the variations, as the spectral resolution of the NEID data is too low to measure line profile changes induced by granulation at this level \citep{Palumbo2024}.

\section{Discussion}

\subsection{Mass-inclination degeneracy}

Given that the orbital inclination of \planet\ is not known, our RV measurements are sufficient only to measure the projected mass, $m_p\sin\,i$, not the true mass. Still, we can determine the likelihood that this object is a planet and not a brown dwarf or even a stellar companion. Assuming that the orbit is drawn from an isotropic distribution uniform in $\cos\,i$, we find that there is only a $0.04\%$ chance that the true mass is $m_p>{\rm M}_{\rm J}$ and a $0.0003\%$ chance that the true mass exceeds $10\,{\rm M}_{\rm J}$. We thus confidently call \planet\ a planet.

\subsection{Comparison to candidate signal}
\citet{Hirsch2021} report an orbital period an eccentricity of $P=31.15^{+0.02}_{-0.02}$ and $e=0.2^{+0.2}_{-0.1}$ and RV semi-amplitude of $K=2.0^{+0.3}_{-0.3}$ for the \targetstar\ planet candidate signal they detect. These values are consistent with our results to better than $1\sigma$. Further, the eccentricities measured by both studies are consistent with a circular orbit when accounting for the Lucy-Sweeney bias \citep{Lucy1971}. 

Not only do we see no periodic signals in the Ca II H \& K indicator from the NEID spectra \autoref{fig:periodograms}, but we also find no strong evidence for a 32-day signal in the HIRES $S_{\rm HK}$ data provided in the CLS data release (\autoref{fig:cps_periodograms}). This non-detection supports our conclusion that the RV signal is indeed attributable to a planet, but it is worth noting that it is inconsistent with the \citet{Hirsch2021} result.

\subsection{Sensitivity to additional planets}

\begin{figure} 
    \centering
    \includegraphics[width=\linewidth]{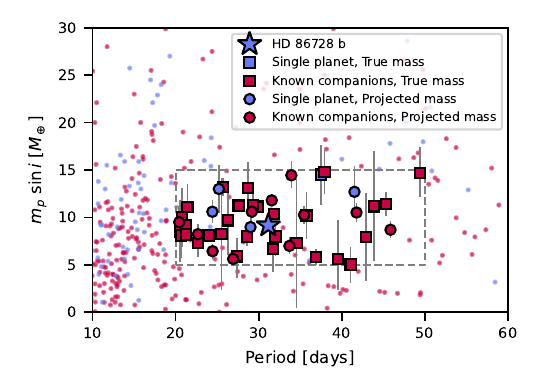}
    \caption{Masses and periods of exoplanets orbiting FGK dwarfs with (red) and without (blue) other known planetary companions in the same system. The grey dashed box highlights exoplanets with masses and periods similar to \planet\ ($5\ $M$_\oplus < m_p\sin i < 15\ $M$_\oplus$; $20$ d $< P < 50 $ d). We differentiate between planets for which the true mass is known (e.g., transiting planets) and planets for which there exists a mass-inclination degeneracy. The former are shown as squares and the latter as circles. \planet\ is in the minority of planets in this region without known siblings. The small red and blue markers indicate planets that fall outside the defined mass-period space or that orbit non-FGK host stars.}
    \label{fig:parameter_space}
\end{figure}

\planet\ joins a larger sample \citep[\autoref{fig:parameter_space}; data taken from][]{ps} of similar mass ($5\ $M$_\oplus < m_p\sin i < 15\ $M$_\oplus$) and period ($20$ d $< P < 50 $ d) planets orbiting FGK dwarfs (4000 K $< T_{\rm eff} < 6500$ K).
Analyses of the intrinsic exoplanet multiplicity distribution from the \textit{Kepler} survey \citep{Borucki2010} reveal that multi-planet systems are common \citep{Lissauer2011,Fang2012,Zink2019,He2019,He2020}; this result has been extended to planetary systems discovered via blind RV searches as well \citep[e.g.,][]{Zhu2022}. The vast majority of the exoplanets in the mass-period space highlighted in \autoref{fig:parameter_space} reside in multi-planet systems. If \targetstar\ has no other planetary companions, \planet\ would be in the small minority ($<12\%$) with no known siblings. Here, we consider our constraints on the presence of additional companions in the system.

To place constraints on our current sensitivity to additional exoplanets, we follow the approach of \citet{Gupta2024} and calculate the expected Fisher information content \citep{Fisher1922} of the existing RV data. We choose this approach rather than an injection-recovery analysis \citep[e.g.,][]{Rosenthal2021} in the interest of computational efficiency, and we note that \citet{Gupta2024} have shown that these two methods produce similar results for high significance signals ($K/\sigma_K\geq5$).

We assume a two-planet model, with a full Keplerian for \planet\ and a simple, zero-eccentricity model for a hypothetical undetected companion
\begin{equation}\label{eq:fisher_model}
    \begin{split}
    \mu(t) = &\ {\rm RV}_{\rm b}(t;K_{\rm b},P_{\rm b},\phi_{\rm b},e_{\rm b},\omega_{\rm b})\ +\\
    &\ {\rm RV}_{\rm c}(t;K_{\rm c},P_{\rm c},\phi_{\rm c})\ +\ \gamma,
    \end{split}
\end{equation}
where $\phi$ is an initial phase offset, $t$ is the set of observation times with length $N = N_{\rm obs,NEID} + N_{\rm obs,CLS}$, and $\gamma$ is an instrument-dependent velocity offset that we set to be equal to either $\gamma_{\rm Lick}$, $\gamma_{\rm HIRES}$, $\gamma_{\rm APF}$, $\gamma_{\rm NEID, Run1}$, or $\gamma_{\rm NEID, Run2}$ using a heaviside function. We calculate the Fisher information matrix as
\begin{equation}\label{eq:fisher_nets_i}
    B_{i,j} = \left(\frac{\partial \mu}{\partial \theta_i}\right)^TC^{-1}\left(\frac{\partial \mu}{\partial \theta_j}\right).
\end{equation}
where the parameter vector $\theta$ is
\begin{equation}
    \begin{split}
    \theta = \{&K_{\rm b}, P_{\rm b}, \phi_{\rm b}, e_{\rm b},\omega_b{\rm b},K_{\rm c},P_{\rm c},\phi_{\rm c},\gamma_{\rm Lick},\\
    &\gamma_{\rm HIRES},\gamma_{\rm APF},\gamma_{\rm NEID,Run1}, \gamma_{\rm NEID,Run2}\},
    \end{split}
\end{equation}
and $C$ is the $N\times N$ covariance matrix. Given that we did not find evidence for correlated noise signals on the timescales captured by our observations, a white noise covariance model is assumed. $C$ becomes a diagonal matrix with elements $C_{n,n} = \sigma_{RV,n}^2+\sigma_{\rm inst,n}^2$, where $\sigma_{\rm RV,n}$ is the measurement precision of the observation at time $t_n$ and $\sigma_{\rm inst,n}$ is the appropriate instrument-specific jitter term as given in \autoref{tab:HD86728_fit_params_scalpels} for the NEID + CLS orbit fit. From the Fisher information matrix, we can calculate the precision on each orbital parameter as
\begin{equation}\label{eq:fisher_sig_i}
    \sigma_{\theta_i}^2 = B_{i,i}^{-1}.
\end{equation}

\begin{figure} 
    \centering
    \includegraphics[width=\linewidth]{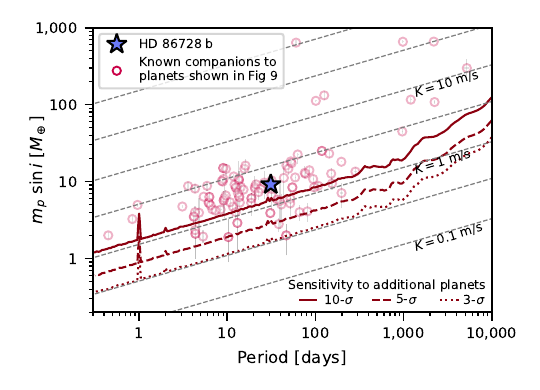}
    \caption{Sensitivity of the existing NEID and CLS RVs to additional exoplanets in the \targetstar\ system, as calculated via Fisher information. The solid, dashed, and dotted red lines show the limits at which the RV semi-amplitude could be measured to 10-, 5-, and 3-$\sigma$ precision, respectively. We note that these limits are calculated assuming white noise variability dominates the stellar RV signal for the observed timescales and cadences, in accordance with our adopted planet fit. Known companions to exoplanets highlighted in \autoref{fig:parameter_space} are shown as open red circles; most of these lie above the expected 10-$\sigma$ sensitivity limits, and all lie above the 3-$\sigma$ limit.}
    \label{fig:hd86728_fisherk}
\end{figure}

We calculate the expected precision on $K_c$ for planets with  $0.3$ d $\leq P_{\rm c} \leq$ $10000$ d and $0.3$\ M$_\oplus\leq m_{\rm c}\sin i \leq 1000\ $M$_\oplus$, marginalizing over orbital phase $\phi_{\rm c}$. The remaining parameters are all set to the best-fit values from our adopted solution in \autoref{tab:HD86728_fit_params_scalpels}.
We show the expected detection sensitivity, $K/\sigma_K$, as a function of $P$ and $m_p\sin i$ in \autoref{fig:hd86728_fisherk}. The existing RV data rule out Jupiter analogs as well as planets as small as $m_p\sin i=$5\ M$_\oplus$ out to orbital periods of $P=100$ d at the 5-$\sigma$ level and out to periods of $P\approx1$ yr at the 3-$\sigma$ level. Planetary companions with lower masses and/or longer periods may still be hidden beneath the noise, both for \targetstar\ as well as for the hosts of the other planets shown in \autoref{fig:parameter_space} and \autoref{fig:hd86728_fisherk}. But as the brightest of these stars, and with intrinsic variability of just $\sim1$m~s$^{-1}$, \targetstar\ offers the best chance to continue to beat down the noise floor.

\section{Conclusion}

We precisely measure the mass and orbit of \planet, a $9\ \rm{M}_\oplus$, 31 d exoplanet, confirming the candidate first identified by \citet{Hirsch2021}. Our analysis of new data from the NEID Earth Twin Survey yields no significant power in any stellar activity indicators at or near the putative orbital period and the RVs provide strong evidence supporting the planetary interpretation. While we find weak activity signals at periods close to 1 year, we show that these are likely due to telluric contamination, which does not affect the RVs. Excess scatter in the residuals to the fit can most likely be attributed to short-timescale stellar variability such as granulation or supergranulation.

\planet\ stands out from most exoplanets with similar masses and periods in that no other planets are detected in the system. The existing RV data are sufficient to rule out additional planetary companions with semi-amplitudes $>1$ m~s$^{-1}$, but planets with lower masses or longer periods may still be hidden in the noise.

The strength of the exoplanet signal seen in the NEID data as compared to archival RVs shows that NEID is delivering on the promise of its exquisite measurement precision. Further, we demonstrate that with the addition of the NEID RVs, we are now sensitive to signals at the sub-m~s$^{-1}$ level for \targetstar. Stellar variability at this level still demands a satisfactory mitigation approach \citep{Crass2021,Luhn2023}. But NEID and other extreme precision spectrographs will continue to chip away at the sensitivity floor for quiet stars, enabling the detection of exoplanets at lower masses and longer periods as we push towards the discovery of Earth analogs with RV measurements.

\section{Acknowledgements}

NEID is funded by NASA through JPL by contract 1547612 and the NEID Data Reduction Pipeline is funded through JPL contract 1644767. Funding for this work was partially provided by Research Support Agreements 1646897 and 1679618 administered by JPL.
The Center for Exoplanets and Habitable Worlds and the Penn State Extraterrestrial Intelligence Center are supported by the Pennsylvania State University and the Eberly College of Science.
This research has made use of the SIMBAD database, operated at CDS, Strasbourg, France, and NASA's Astrophysics Data System Bibliographic Services.
Part of this work was performed at the Jet Propulsion Laboratory, California Institute of Technology, sponsored by the United States Government under the Prime Contract 80NM0018D0004 between Caltech and NASA. 

This paper contains data taken with the NEID instrument, which was funded by the NASA-NSF Exoplanet Observational Research (NN-EXPLORE) partnership and built by Pennsylvania State University. NEID is installed on the WIYN telescope, which is operated by the NSF's National Optical-Infrared Astronomy Research Laboratory, and the
NEID archive is operated by the NASA Exoplanet Science Institute at the California Institute of Technology.
NN-EXPLORE is managed by the Jet Propulsion Laboratory, California Institute of Technology under contract with the National Aeronautics and Space Administration. We thank the NEID Queue Observers and WIYN Observing Associates for their skillful execution of our observations.
CIC acknowledges support by NASA Headquarters through an appointment to the NASA Postdoctoral Program at the Goddard Space Flight Center, administered by ORAU through a contract with NASA.

This research made use of \texttt{exoplanet} \citep{Foreman-Mackey2021,exoplanet:zenodo} and its dependencies \citep{Foreman-Mackey2017,Foreman-Mackey2018,Kumar2019,AstropyCollaboration2018,Salvatier2016}.
Based in part on observations at Kitt Peak National Observatory, NSF’s NOIRLab, managed by the Association of Universities for Research in Astronomy (AURA) under a cooperative agreement with the National Science Foundation. The authors are honored to be permitted to conduct astronomical research on Iolkam Du’ag (Kitt Peak), a mountain with particular significance to the Tohono O’odham.
We also express our deepest gratitude to Zade Arnold, Joe Davis, Michelle Edwards, John Ehret, Tina Juan, Brian Pisarek, Aaron Rowe, Fred Wortman, the Eastern Area Incident Management Team, and all of the firefighters and air support crew who fought the recent Contreras fire. Against great odds, you saved Kitt Peak National Observatory.

The Pennsylvania State University campuses are located on the original homelands of the Erie, Haudenosaunee (Seneca, Cayuga, Onondaga, Oneida, Mohawk, and Tuscarora), Lenape (Delaware Nation, Delaware Tribe, Stockbridge-Munsee), Shawnee (Absentee, Eastern, and Oklahoma), Susquehannock, and Wahzhazhe (Osage) Nations.  As a land grant institution, we acknowledge and honor the traditional caretakers of these lands and strive to understand and model their responsible stewardship. We also acknowledge the longer history of these lands and our place in that history.

\facilities{WIYN (NEID)}

\software{\texttt{Agatha} \citep{Feng2017}, \texttt{astropy} \citep{AstropyCollaboration2018}, \texttt{arviz} \citep{Kumar2019}, \texttt{barycorrpy} \citep{Kanodia2018}, \texttt{lightkurve}\citep{LightkurveCollaboration2018}, \texttt{matplotlib} \citep{Hunter2007}, \texttt{numpy} \citep{Harris2020}, \texttt{PyMC3} \citep{Salvatier2016}, \texttt{scipy} \citep{Oliphant2007}, \texttt{Theano} \citep{TheTheanoDevelopmentTeam2016}}

\bibliography{references}{}
\bibliographystyle{aasjournal}

\appendix

\section{NEID Wavelength Offset}

NEID was thermally cycled to protect the spectrograph from the Contreras fire in 2022. Differences in the construction of the wavelength solution pre- and post-fire result in a RV zero point offset, which must be accounted for when comparing data taken on both sides of the intervention. We denote the velocity measured by the NEID DRP in order $j$ at observation time $t_i$ as $\varv_{i,j}$ and the measurement uncertainty as $\sigma_{i,j}$. The full RV time series in each order is $\varv_j$, and the time series of measurements within a given run is $\varv_{i\in U_u,j}$. Here, $U_u$ is the union of all times within a run $u$, where the time series is separated into distinct runs at known velocity breaks. For the data set in this work, $u=1$ corresponds to data taken during Run 1, before the velocity break following the Contreras fire, and $u=2$ corresponds to data taken after the fire.

First, we use the CCF modules from the DRP to calculate a velocity, $\varv_{i,{\rm LFC}}$, and measurement uncertainty, $\sigma_{i,{\rm LFC}}$, for each observation using only the orders tied to the LFC wavelength solution ($63\leq j \leq 118$). We use these values to estimate the average LFC velocity offset for each run as the difference between the mean during run $u$ and the mean of the entire time series: 
\begin{equation}
    \Delta_{u,{\rm LFC}} = \left(\sum_{i\in U_u} \frac{\varv_{i,{\rm LFC}}}{\sigma_{i,{\rm LFC}}^{2}} \middle/ \sum_{i\in U_u}\frac{1}{\sigma_{i,{\rm LFC}}^{2}}\right) - \left(\sum_{i} \frac{\varv_{i,{\rm LFC}}}{\sigma_{i,{\rm LFC}}^{2}} \middle/ \sum_{i}\frac{1}{\sigma_{i,{\rm LFC}}^{2}}\right),
\end{equation}
where the velocities are weighted by the inverse variance, $\sigma_{i,{\rm LFC}}^{-2}$.

We then calculate initial order-by-order velocity offsets for each run in the same manner:
\begin{equation}
    \Delta^0_{u,j} = \left(\sum_{i\in U_u} \frac{\varv_{i,j}}{\sigma_{i,j}^{2}} \middle/ \sum_{i\in U_u}\frac{1}{\sigma_{i,j}^{2}}\right) - \left(\sum_{i} \frac{\varv_{i,j}}{\sigma_{i,j}^{2}} \middle/ \sum_{i}\frac{1}{\sigma_{i,j}^{2}}\right).
\end{equation}
However, these order-dependent values will contain not only the offset introduced by the aforementioned pipeline bug, but also any real astrophysical variation between different runs. Taking $\Delta_{u,{\rm LFC}}$ to be a good estimate of the actual velocity change, we subtract it from $\Delta^0_{u,j}$ to preserve the signal:
\begin{equation}
    \Delta^1_{u,j} = \Delta^0_{u,j} - \Delta_{u,{\rm LFC}}
\end{equation}

Finally, we shift the CCF of each order by$\Delta^1_{u,j}$ then use the DRP modules to compute new weighted velocities and measurement uncertainties for each observation. 

\end{document}